\documentclass[aps,preprint]{revtex4}
\usepackage{amssymb}
\usepackage{amsmath}

\input{tcilatex}

\begin{document}

\title{Stringy Symmetries and Their High-energy Limits}
\author{Chuan-Tsung Chan}
\email{ctchan@phys.cts.nthu.edu.tw}
\affiliation{Department of International Trade, Lan Yang Institute of Technology, I-Lan,
Taiwan, R.O.C.}
\author{Jen-Chi Lee}
\email{jcclee@cc.nctu.edu.tw}
\affiliation{Department of Electrophysics, National Chiao-Tung University, Hsinchu,
Taiwan, R.O.C.}
\date{\today }

\begin{abstract}
We derive stringy symmetries with conserved charges of arbitrarily high
spins from the decoupling of two types of zero-norm states in the old
covariant first quantized (OCFQ) spectrum of open bosonic string. These
symmetries are valid to all energy $\alpha ^{\prime }$ and all loop orders $%
\chi $ in string perturbation theory. The high-energy limit $\alpha ^{\prime
}\rightarrow \infty $ of these stringy symmetries can then be used to fix
the proportionality constants between scattering amplitudes of different
string states algebraically \textit{without} referring to Gross and Mende's
saddle point calculation of high-energy string-loop amplitudes. These
proportionality constants are, as conjectured by Gross, independent of the
scattering angle $\phi _{CM}$ and the order $\chi $ of string perturbation
theory. However, we also discover some \textit{new} nonzero components of
high-energy amplitudes not found previously by Gross and Manes. These
components are essential to preserve massive gauge invariances or decouple
massive zero-norm states of string theory. A set of massive scattering
amplitudes and their high energy limit are calculated explicitly to justify
our results.
\end{abstract}

\maketitle

%\author{Chuan-Tsung Chan\thanks{email address: ctchan@ep.nctu.edu.tw} \
%         and Jen-Chi Lee\thanks{email address: jcclee@cc.nctu.edu.tw}}
%\email{ctchan@ep.nctu.edu.tw and jcclee@cc.nctu.edu.tw}
%\affiliation{Department of International Trade, Lan Yang Institute
%of Technology, I-Lan, Taiwan, R.O.C.\\ and
%\\Department of Electrophysics, National Chiao-Tung
%University, Hsinchu, Taiwan, R.O.C.}

In the traditional formulation of a local quantum field theory, a symmetry
principle was postulated, which can be used to determine the interaction of
the theory, e.g., Yang-Mills theories and general relativity. The idea of
\textquotedblleft\ symmetry dictates interaction\textquotedblright\ has thus
become one of the fundamental philosophy to pursue new physics such as GUTs
and supergravities for the last few decades. One of the most important
consequences of these symmetries is the resulting softer ultraviolet
structure of field theories which, in some cases, makes them consistent or
renormalizable quantum field theories when incorporating with quantum
mechanics. In these cases, the Ward identities, the direct consequence of
symmetry on the n-point Green functions of the theory, are intensively used
to remove the unwanted loop divergences in perturbation theory. In contrast
to the local quantum field theory, string theory is very different in this
respect. In string theory, on the contrary, it is the interaction,
prescribed by the very tight quantum consistency conditions due to the
extendedness of string rather than point particle, which determines the form
of the symmetry. In fact, once we settle on the quantum theory of a free
string, the forms of the interactions and thus symmetries of all string
states are fixed by the quantum consistency of the theory. For example, the
massless gauge symmetries of 10D Heterotic string\cite{1} were determined to
be SO(32) or $E_{8}^{2}$ by the string one-loop consistency or modular
invariance of the theory. Some stringy Einstein-Yang-Mills type symmetries
with symmetry parameters containing both Einstein and Yang-Mills index were
proposed in Ref[2]. Being a consistent quantum theory with no free parameter
and an infinite number of states, it is conceivable that there exists an
huge symmetry group or Ward identities, which are responsible for the
ultraviolet finiteness of string theory. To uncover the structure of this
huge hidden symmetry group has become one of the most challenging problem
ever since the discovery of string theory.

In 1988 Gross\cite{3} made an important progress on this subject (see also 
\cite{4} for the subsequent developments). With the calculation of
high-energy limit of closed string scattering amplitudes for an arbitrary
string-loop order G through the use of a semi-classical, saddle point
technique developed by Gross and Mende\cite{5}, he was able to derive an
infinite number of linear relations among high-energy scattering amplitudes
of different string states with the same momenta. These relations were shown
to be valid order by order and were of the \textit{identical form} in string
perturbation theory. As a result, the high-energy scattering amplitudes of
all string states can be expressed in terms of, say, the dilaton scattering
amplitudes. A similar result was obtained for the open string by Gross and
Manes\cite{6}. However, the physical origin of these symmetries and thus the
meaning of proportionality constants between the high-energy scattering
amplitudes of different string states were unknown to those authors, and
their values were not calculated.

In this letter, we propose an infinite number of stringy Ward identities
derived from the decoupling of two types of zero-norm states\cite{7} in the
OCFQ string spectrum. These Ward identities are valid to \textit{all} energy 
$\alpha ^{\prime }$ and to all loop orders in string perturbation theory
since zero-norm states should be decoupled from the correlation functions at
each order of perturbation theory by unitarity. The simplest example is the
familiar massless \textit{on-shell} Ward identity of string QED. In this
sense, the stringy Ward identities we proposed in this letter serve as a
natural generalization of Ward identity in gauge field theory. As the first
test of these stringy Ward identities, the high-energy limit $\alpha
^{\prime }\rightarrow \infty $ of them are used to produce Gross's \cite{3}
linear relations among high-energy scattering amplitudes of different string
states with the same momenta. Moreover, the proportionality constants
between scattering amplitudes of different string states are calculated for
the second massive level algebraically \textit{without} referring to Gross
and Mende's saddle point calculation of high-energy string-loop amplitudes.
Our calculation thus serves as a consistent check of the saddle point
technique of string-loop diagram developed by Gross and Mende \cite{5}. We
find that these high-energy proportionality constants are, as conjectured by
Gross \cite{3}, independent of scattering angle $\phi _{CM}$ and the order $%
\chi $ of string perturbation theory. However, the proportionality
coefficients do depend on the scattering angle $\phi _{CM}$ through the
dependence of momentum at \textit{low} energy. \textit{More importantly, we
also discover some new nonzero components of high-energy amplitudes not
found previously by Gross and Manes\cite{6}. These components are essential
to preserve massive gauge invariances or decouple massive zero-norm states
of string theory. }As an explicit example, we calculate the high energy
limit of a set of massive scattering amplitudes of the second massive level
derived in \cite{8} to justify our results. The fact that zero-norm states
imply inter-particle symmetries was demonstrated previously by two other
independent approaches based on the massive worldsheet sigma-model\cite{9}
and Witten's string field theory\cite{10}. To further uncover the group
theoretical structure of these stringy symmetries, it is important to
explicitly calculate the complete set of zero-norm states with arbitrarily
high spins in the spectrum. Recently, a simplified method to generate
zero-norm states in OCFQ bosonic string was proposed\cite{11}. General
formulas of some zero-norm tensor states at an arbitrary mass level were
given. Unfortunately, general formulas for the \textit{complete} set of
zero-norm states are still lacking mostly due to the high dimensionality of
spacetime D = 26. However, in the toy 2D string model\cite{12}, a general
formula of zero-norm states with discrete Polyakov's momenta at an arbitrary
mass level was given in terms of Schur Polynomials\cite{13}. These zero-norm
states were shown to carry the spacetime $\omega _{\infty }$ charges. On the
other hand, the complete spacetime symmetry group of toy 2D string was known
to be the same $\omega _{\infty }$, and the corresponding $\omega _{\infty }$
Ward identities were powerful enough to determine the tachyon scattering
amplitudes \textit{without }any integration. These observations in 2D and
26D string theories signal the importance of the existence of zero-norm
states in the OCFQ string spectrum, not shared by other quantization schemes
of string theory, e.g., light-cone quantization. The advantage of using the
decoupling of zero-norm states to derive stringy Ward identities is that one
can avoid the difficult calculation of string- loop amplitudes. Another one
is that the resulting Ward identities are valid to \textit{all} energy $%
\alpha ^{\prime }$, in contrast to the high-energy $\alpha ^{\prime
}\rightarrow \infty $ result of Gross.

Let's begin with a brief review of QED Ward identity

\begin{equation}
k_{\mu _{1}}\mathcal{T}^{\mu _{1}\mu _{2}\cdots \mu _{n}}(k_{1}k_{2}\cdots
k_{n})=0,
\end{equation}
where $\mathcal{T}$ is the\textit{\ off-shell} n-point Green function for n
external photons of polarizations $\mu _{1},\cdots ,\mu _{n}$ and momenta $%
k_{1},\cdots ,k_{n}.$ eq.(1) means that the amplitude $\mathcal{T}$\emph{\ }
vanishes if the polarization of one of the external photons is taken to be
longitudinal. Note that eq. (1) holds even off-shell. This seemingly simple
equation, which originated from U(1) gauge symmetry, turns out to be one of
the most far-reaching property of QED. In the old covariant Gupta-Bleuler
quantization of QED, the polarization vector $\epsilon _{\mu}$ of photon is
constrained by the covariant gauge condition $%
%TCIMACRO{\FORMULA{k\cdot \varepsilon =0}{k\cdot \varepsilon =0}{evaluate}}%
%BeginExpansion
k\cdot \varepsilon =0%
%EndExpansion
$. One of the three allowed physical polarizations, the longitudinal one $%
%TCIMACRO{\FORMULA{\varepsilon =k}{\varepsilon =k}{evaluate}}%
%BeginExpansion
\varepsilon =k%
%EndExpansion
$, is zero-norm due to the massless condition of on-shell photon. The theory
thus ends up with only two physical transverse propagating modes, and the
longitudinal degree of freedom turns out to serve as the U(1) symmetry
parameter of the theory. In the OCFQ spectrum of open bosonic string theory,
there exists a natural stringy generalization of this zero-norm longitudinal
degree of freedom. They are(we use the notation in Ref[7])

\begin{equation}
\text{Type I}:L_{-1}\left| x\right\rangle ,\text{ where }L_{1}\left|
x\right\rangle =L_{2}\left| x\right\rangle =0,\text{ }L_{0}\left|
x\right\rangle =0;
\end{equation}

\begin{equation}
\text{Type II}:(L_{-2}+\frac{3}{2}L_{-1}^{2})\left\vert \widetilde{x}%
\right\rangle ,\text{ where }L_{1}\left\vert \widetilde{x}\right\rangle
=L_{2}\left\vert \widetilde{x}\right\rangle =0,\text{ }(L_{0}+1)\left\vert 
\widetilde{x}\right\rangle =0.
\end{equation}
While type I states have zero-norm at any space-time dimension,type II
states have zero-norm \emph{only} at D=26. The existence of type II
zero-norm states turns out to be crucial for the discussion in the rest of
this letter. The simplest zero-norm state $k\cdot \alpha _{-1}\mid
0,k\rangle $, $%
%TCIMACRO{\FORMULA{k^{2}=0}{k^{2}=0}{evaluate}}%
%BeginExpansion
k^{2}=0%
%EndExpansion
$ with polarization $k$ is the massless solution of eq. (2), which
reproduces the longitudinal photon discussed in eq. (1). A simple
prescription to systematically solve eqs. (2) and (3) for an infinite number
of zero-norm states was given recently in Ref[11]. A more thorough
understanding of the solution of these equations and their relation to
space-time $\omega _{\infty }$ symmetry of toy D=2 string was discussed in
Ref[13]. \ 

In the first quantized approach of string theory, the string generalization
of eq(1), or the stringy \textit{on-shell} Ward identities are proposed to
be (for our purpose we choose four-point amplitudes in this letter)

\begin{equation}
\mathcal{T}_{\chi }(k_{i})=g_{c}^{2-\chi }\int \frac{Dg_{\alpha \beta }}{%
\mathcal{N}}DX^{\mu }\exp (-\frac{\alpha ^{\prime }}{2\pi }\int d^{2}\xi 
\sqrt{g}g^{\alpha \beta }\partial _{\alpha }X^{\mu }\partial _{\beta }X_{\mu
})\overset{4}{\underset{i=1}{\Pi }}v_{i}(k_{i})=0,
\end{equation}%
where at least one of the 4 vertex operators corresponds to the zero-norm
state solution of eqs. (2) or (3). In eq(4) $g_{c}$ is the closed string
coupling constant, $\mathcal{N}$ is the volume of the group of
diffeomorphisms and Weyl rescalings of the worldsheet metric, and $%
v_{i}(k_{i})$ are the on-shell vertex operators with momenta $k_{i}$. The
integral is over orientable open surfaces of Euler number $\chi $
parametrized by moduli $\overrightarrow{m}$ with punctures at $\xi _{i}$. To
illustrate the power of this seemingly trivial equation, the four Ward
identities of the second massive level (spin-three) were calculated to be%
\cite{8}

\begin{equation}
k_{\mu }\theta _{\nu \lambda }\mathcal{T}_{\chi }^{(\mu \nu \lambda
)}+2\theta _{\mu \nu }\mathcal{T}_{\chi }^{(\mu \nu )}=0,
\end{equation}%
\begin{equation}
(\frac{5}{2}k_{\mu }k_{\nu }\theta _{\lambda }^{\prime }+\eta _{\mu \nu
}\theta _{\lambda }^{\prime })\mathcal{T}_{\chi }^{(\mu \nu \lambda
)}+9k_{\mu }\theta _{\nu }^{\prime }\mathcal{T}_{\chi }^{(\mu \nu )}+6\theta
_{\mu }^{\prime }\mathcal{T}_{\chi }^{\mu }=0,
\end{equation}%
\begin{equation}
(\frac{1}{2}k_{\mu }k_{\nu }\theta _{\lambda }+2\eta _{\mu \nu }\theta
_{\lambda })\mathcal{T}_{\chi }^{(\mu \nu \lambda )}+9k_{\mu }\theta _{\nu }%
\mathcal{T}_{\chi }^{[\mu \nu ]}-6\theta _{\mu }\mathcal{T}_{\chi }^{\mu }=0,
\end{equation}%
\begin{equation}
(\frac{17}{4}k_{\mu }k_{\nu }k_{\lambda }+\frac{9}{2}\eta _{\mu \nu
}k_{\lambda })\mathcal{T}_{\chi }^{(\mu \nu \lambda )}+(9\eta _{\mu \nu
}+21k_{\mu }k_{\nu })\mathcal{T}_{\chi }^{(\mu \nu )}+25k_{\mu }\mathcal{T}%
_{\chi }^{\mu }=0,
\end{equation}%
where $\theta _{\mu \nu }$ is transverse and traceless, and $\theta
_{\lambda }^{\prime }$ and $\theta _{\lambda }$ are transverse vectors. In
each equation, we have chosen, say, $v_{2}(k_{2})$\ to be the vertex
operators constructed from zero-norm states and $k_{\mu }\equiv k_{2\mu }$.
Note that eq.(7) is the inter-particle Ward identity corresponding to $D_{2}$
vector zero-norm state obtained by antisymmetrizing those terms which
contain $\alpha _{-1}^{\mu }\alpha _{-2}^{\nu }$ in the original type I and
type II vector zero-norm states. We will use 1 and 2 for the incoming
particles and 3 and 4 for the scattered particles. In eqs. (5)-(8), 1,3 and
4 can be any string states (including zero-norm states) and we have omitted
their tensor indices for the cases of excited string states. For example,
one can choose $v_{1}(k_{1})$\ to be the vertex operator constructed from
another zero-norm state which generates an inter-particle Ward identity of
the third massive level. The resulting Ward-identity of eq (7) then relates
scattering amplitudes of particles at different mass level. $\mathcal{T}%
_{\chi }^{\prime }s$ in eqs (5)-(8) are the second massive level, $\chi $-th
order string-loop amplitudes. For the string-tree level $\chi $=1 with three
tachyons $v_{1,3,4}$, the three scattering amplitudes $\mathcal{T}_{\chi
}^{\prime }s$ were explicitly calculated and the Ward identities eqs(5)-(8)
were verified \cite{8}. At this point, \{$\mathcal{T}_{\chi }^{(\mu \nu
\lambda )},\mathcal{T}_{\chi }^{(\mu \nu )},\mathcal{T}_{\chi }^{\mu }$\} is
identified to be the \emph{amplitude triplet} of the spin-three state. In
fact, it can be shown that $\mathcal{T}_{\chi }^{(\mu \nu )}$ and $\mathcal{T%
}_{\chi }^{\mu }$ are fixed by $\mathcal{T}_{\chi }^{(\mu \nu \lambda )}$
due to the stringy Ward identities, eqs.(5) and (6), constructed from the
type I spin-two zero-norm state and another vector zero-norm state obtained
by symmetrizing those terms which contain $\alpha _{-1}^{\mu }\alpha
_{-2}^{\nu }$ in the original type I and type II vector zero-norm states. $%
\mathcal{T}_{\chi }^{[\mu \nu ]}$ is obviously identified to be the
scattering amplitude of the antisymmetric spin-two state with the same
momenta as $\mathcal{T}_{\chi }^{(\mu \nu \lambda )}$.Eq. (7) thus relates
the scattering amplitudes of two different string states at the second
massive level. Note that eqs. (5)-(8) are valid order by order and are \emph{%
automatically} of the identical form in string perturbation theory. This is
consistent with Gross's argument through the calculation of high-energy
scattering amplitudes. However, it is important to note that eqs. (5)-(8)
are, in contrast to the high-energy $\alpha ^{\prime }\rightarrow \infty $
result of Gross, valid to \emph{all} energy $\alpha ^{\prime }$ and their
coefficients do depend on the center of mass scattering angle $\phi _{CM}$ ,
which is defined to be the angle between $\overrightarrow{k}_{1}$ and $%
\overrightarrow{k}_{3}$, through the dependence of momentum $k$ . To produce
Gross's high-energy result and fix the proportionality constants, which were
not dwelt on in Ref[3,6] due to lack of the physical origin of the proposed
high-energy symmetries, one needs to calculate high-energy limit of eqs.
(5)-(8).

\bigskip We will calculate high energy limit of eqs.(5)-(8) without
referring to the saddle point calculation in \cite{3,5,6}. Let's define the
normalized polarization vectors,$e_{P}=\frac{1}{m_{2}}(E_{2},\mathrm{k}%
_{2},0) = \frac{k_{2}}{m_{2}},$ $e_{L}= \frac{1}{m_{2}}(\mathrm{k}%
_{2},E_{2},0)$ and $e_{T}$ $=(0,0,1)$ in the CM frame contained in the plane
of scattering. They satisfy the completeness relation $\eta ^{\mu \nu
}=\Sigma _{\alpha ,\beta }e_{\alpha }^{\mu }e_{\beta }^{\nu }$ $\eta
^{\alpha \beta }$, where $\mu ,\nu =0,1,2$ and $\alpha ,\beta =P,L,T.$ $Diag$
$\eta ^{\mu \nu }=(-1,1,1).$One can now transform all $\mu ,\nu $
coordinates in eqs.(5)-(8) to coordinates $\alpha ,\beta$. For eq(5), we
have $\theta^{\mu \nu}= e_{L}^{\mu } e_{L}^{\nu}-e_{T}^{\mu }e_{T}^{\nu }$
or $\theta^{\mu \nu} = e_{L}^{\mu }e_{T}^{\nu }+ e_{T}^{\mu }e_{L}^{\nu }$ .
In the high energy $E\rightarrow $ $\infty ,$ fixed angle $\phi _{CM}$
limit, one identifies $e_{P}=e_{L}$ and eq. (5) gives ( we drop loop order $%
\chi $ here to simplify the notation)

\begin{equation}
\mathcal{T}_{LLL}^{6\rightarrow 4}-\mathcal{T}_{LTT}^{4}+\mathcal{T}%
_{(LL)}^{4}-\mathcal{T}_{(TT)}^{2}=0,
\end{equation}%
\begin{equation}
\mathcal{T}_{LLT}^{5\rightarrow 3}+\mathcal{T}_{(LT)}^{3}=0.
\end{equation}%
In eqs (9) and (10), we have assigned a relative energy power for each
amplitude. For each $L$ component, the order is $E^{2}$ ( the naive order of 
$e_{L}\cdot k$ is $E^{2}$ ) and for each transverse $T$ component, the order
is $E$ ( the naive order of $e_{T}\cdot k$ is $E$ )$.$ This is due to the
definitions of $\ e_{L}$ and $e_{T}$ above, where $e_{L}$ got one energy
power more than $e_{T}.$ Thus, for example, the naive order of $\mathcal{T}%
_{LLL}$ is $E^{6}$. However, by eq. (9), the $E^{6}$ term of the energy
expansion for $\mathcal{T}_{LLL}$ is forced to be zero. As a result, the
leading order term of $\mathcal{T}_{LLL}$ is at most $E^{4}$. We have used $%
6\rightarrow 4$ in eq.(9) to represent this energy reduction. Similar rule
applies to $\mathcal{T}_{LLT}$ in eq(10). For eq(6), we have $\theta
^{\prime \mu }=e_{L}^{\mu }$ or $\theta ^{\prime \mu }=e_{T}^{\mu }$ and one
gets, in the high energy limit,

\begin{equation}
10\mathcal{T}_{LLL}^{6\rightarrow 4}+\mathcal{T}_{LTT}^{4}+18\mathcal{T}%
_{(LL)}^{4}+6\mathcal{T}_{L}^{2}=0,
\end{equation}%
\begin{equation}
10\mathcal{T}_{LLT}^{5\rightarrow 3}+\mathcal{T}_{TTT}^{3}+18\mathcal{T}%
_{(LT)}^{3}+6\mathcal{T}_{T}^{1}=0.
\end{equation}%
For the $D_{2}$ Ward identity, eq.(7), we have $\theta ^{\mu }= e_{L}^{\mu }$
or $\theta ^{\mu }=e _{T}^{\mu }$ and one gets, in the high energy limit,

\begin{equation}
\mathcal{T}_{LLL}^{6\rightarrow 4}+\mathcal{T}_{LTT}^{4}+9\mathcal{T}%
_{[LL]}^{4\rightarrow 2}-3\mathcal{T}_{L}^{2}=0,
\end{equation}

\begin{equation}
\mathcal{T}_{LLT}^{5\rightarrow 3}+\mathcal{T}_{TTT}^{3}+9\mathcal{T}%
_{[LT]}^{3}-3\mathcal{T}_{T}^{1}=0.
\end{equation}%
Note that $\mathcal{T}_{[LL]}$ in eq.(13) originate from the high energy
limit of $\mathcal{T}_{[PL]}$, and the antisymmetric property of the tensor
forces the leading $E^{4}$ term to be zero. Finally the singlet zero norm
state Ward identity, eq.(8), imply, in the high energy limit,

\begin{equation}
34\mathcal{T}_{LLL}^{6\rightarrow 4}+9\mathcal{T}_{LTT}^{4}+84\mathcal{T}%
_{(LL)}^{4}+9\mathcal{T}_{(TT)}^{2}+50\mathcal{T}_{L}^{2}=0.
\end{equation}%
It is important to note that all components of high energy amplitudes of
symmetric spin three and antisymmetric spin two states appear at least once
in eqs. (9)-(15). It is now easy to see that the naive leading order
amplitudes corresponding to $E^{4}$ \ appear in eqs.(9), (11), (13) and
(15). However, a simple calculation shows that $\mathcal{T}_{LLL}^{4}=%
\mathcal{T}_{LTT}^{4}=\mathcal{T}_{(LL)}^{4}=0.$So the real leading order
amplitudes correspond to $E^{3}$, which appear in eqs.(10), (12) and (14). A
simple calculation shows that

\begin{equation}
\mathcal{T}_{TTT}^{3}:\mathcal{T}_{LLT}^{3}:\mathcal{T}_{(LT)}^{3}:\mathcal{T%
}_{[LT]}^{3}=8:1:-1:-1.
\end{equation}%
Note that these proportionality constants are, as conjectured by Gross,
independent of the scattering angle $\phi _{CM}$ and the loop order $\chi $
of string perturbation theory. \textit{Most importantly, we now understand
that they originate from zero-norm states in the OCFQ spectrum of the
string! }The subleading order amplitudes corresponding to $E^{2}$appear in
eqs.(9), (11), (13) and (15). One has 6 unknown amplitudes and 4 equations.
Presumably, they are not proportional to each other or the proportional
coefficients do depend on the scattering angle $\phi _{CM}$. We will justify
this point later in our sample calculation. Our calculation here is, similar
to the toy 2D string case, purely algebraic \textit{without any integration}
and is independent of saddle point calculation in \cite{3,5,6}. It is
important to note that our result in eq.(16) is gauge invariant as it should
be since we derive it from Ward identities (5)-(8). On the other hand, the
result obtained in \cite{6} with $\mathcal{T}_{TTT}\propto \mathcal{T}%
_{[LT]},$ and $\mathcal{T}_{LLT}=\mathcal{T}_{(LT)}=0$ in the leading order
energy at this mass level is, on the contrary, \textit{not} gauge invariant.
In fact, with only two non-zer$\Sigma $o amplitudes of $\mathcal{T}_{TTT}$
and $\mathcal{T}_{[LT]}$, an inconsistency arises between eqs. (6) and (7)
or eqs. (12) and (14). To further justify our result, we give a sample
calculation. For the string-tree level $\chi $=1, with one tensor and three
tachyons $v_{1,3,4}$, the four scattering amplitudes$\mathcal{T}^{(\mu \nu
\lambda )},\mathcal{T}^{(\mu \nu )},\mathcal{T}^{[\mu \nu ]}$ and $\mathcal{T%
}^{\mu }$ were explicitly calculated in \cite{8}. An explicit calculation of
their high energy limits give the kinematic factors of the amplitudes ($s-t$
channel only) $\mathcal{K}_{TTT}=-8E^{9}\sin ^{3}\phi _{CM}=8\mathcal{K}%
_{LLT}=-8\mathcal{K}_{(LT)}=-8\mathcal{K}_{[LT]},$where $%
%TCIMACRO{\FORMULA{s=-(k_{1}+k_{2})^{2}}{s=-(k_{1}+k_{2})^{2}}{evaluate}}%
%BeginExpansion
s=-(k_{1}+k_{2})^{2}%
%EndExpansion
$, $%
%TCIMACRO{\FORMULA{t=-(k_{2}+k_{3})^{2}}{t=-(k_{2}+k_{3})^{2}}{evaluate}}%
%BeginExpansion
t=-(k_{2}+k_{3})^{2}%
%EndExpansion
$,and $%
%TCIMACRO{\FORMULA{u=-(k_{1}+k_{3})^{2}}{u=-(k_{1}+k_{3})^{2}}{evaluate}}%
%BeginExpansion
u=-(k_{1}+k_{3})^{2}%
%EndExpansion
$ are the Mandelstam variables. Also $\mathcal{T}_{LLL}^{6}=\mathcal{T}%
_{LLT}^{5}=0$ as claimed above. A calculation of subleading order in $E$
shows that the amplitudes are not proportional to each other or the
proportional coefficients do depend on the scattering angle $\phi _{CM}.$%
Similar calculations can be done for the third massive level. The result is%
\cite{14}

\begin{eqnarray}
\mathcal{T}_{TTTT}^{4} &:&\mathcal{T}_{TTLL}^{4}:\mathcal{T}_{LLLL}^{4}:%
\mathcal{T}_{TTL}^{4}:\mathcal{T}_{LLL}^{4}:\widetilde{\mathcal{T}}%
_{LT,T}^{4}:\widetilde{\mathcal{T}}_{LP,P}^{4}:\mathcal{T}_{LL}^{4}:%
\widetilde{\mathcal{T}}_{LL}^{4}=  \notag \\
16 &:&\frac{4}{3}:\frac{1}{3}:-\frac{4\sqrt{6}}{9}:-\frac{\sqrt{6}}{9}:-%
\frac{2\sqrt{6}}{3}:0:\frac{2}{3}:0
\end{eqnarray}%
where $\mathcal{T}_{\mu \nu \lambda }$, $\widetilde{\mathcal{T}}_{\mu \nu
,\lambda }$ , $\mathcal{T}_{\mu \nu }$and $\widetilde{\mathcal{T}}_{\mu \nu }
$ are amplitudes corresponding to $\alpha _{-1}^{(\mu \nu }\alpha
_{-2}^{\lambda )}$, mixed-symmetric spin three of $\alpha _{-1}^{\mu \nu
}\alpha _{-2}^{\lambda }$, $\alpha _{-2}^{\mu }\alpha _{-2}^{\nu }$ and $%
\alpha _{-1}^{\mu }\alpha _{-3}^{\nu }$ respectively. It is remarkable to
discover that both algebraic and sample calculations give exactly the same
results Eqs.(16) and (17). In general there is only one independent
component of high-energy scattering amplitude at each fixed mass level, and
it can be deduced that 
\begin{equation}
\mathcal{T}_{n_{1}n_{2}n_{3}n_{4}}^{TT...}=[(-2)E^{3}\sin \phi _{CM}]^{N}%
\mathcal{T}(N),
\end{equation}%
where $n_{i}$ is the number of $T$ for the $i$-th particle and $\mathcal{T}%
(N)=\sqrt{\pi }(-1)^{N-1}2^{-N}E^{-1-2N}(\sin \frac{\phi _{CM}}{2}%
)^{-3}(\cos \frac{\phi _{CM}}{2})^{5-2N}\exp (-\frac{s\ln s+t\ln t-(s+t)\ln
(s+t)}{2}),N=\sum n_{i}$. \textit{As a result, all high-energy string
scattering amplitudes can be expressed in terms of those of tachyons}.
Finally, unlike the saddle point calculation, our algebraic approach is very
easy to generalize to closed string case by "doubling the spectrum". In that
case, one has 32 zero norm state at the second massive level. The non-zero
high energy amplitudes can be obtained by doubling eq.(16), which amounts to
16 non-zero components.

\bigskip We conclude that the physical origin of the high-energy symmetries
and the proportionality constants in eq (16) are from the zero-norm states
in the OCFQ spectrum. The most challenging problem remained is the
calculation of algebraic structure of these stringy symmetries derived from
the complete zero-norm state solutions of eqs. (2) and (3) with arbitrarily
high spins. Presumably, it is a complicated 26D generalization of $\omega
_{\infty }$ of the simpler toy 2D string model \cite{13}.

C.T would like to thank Physics Department of National Taiwan University for
computer facilities. J.C would like to thank Physics Departments of National
Taiwan University and Simon-Fraser University, where early part of this work
was completed during my sabbatical visits. C.T's work is supported through a
NSC grant. J.C's work is supported in part by a grant of NSC and a
travelling fund of government of Taiwan.

\end{document}